\documentstyle[12pt]{article}

\begin{document}

\begin{center}
An example of dissipative quantum system: finite differences
for complex Ginibre ensemble
\end{center}
\begin{center}
Maciej M. Duras
\end{center}
\begin{center}
Institute of Physics, Cracow University of Technology, 
ulica Podchor\c{a}\.zych 1, 30-084 Cracow, Poland
\end{center}

\begin{center}
Email: mduras @ riad.usk.pk.edu.pl
\end{center}

\begin{center}
"Quantum dynamics in terms of phase-space distributions"; 
May 22nd, 2000 to May 26th, 2000; 
Max Planck Institute for the Physics of Complex Systems;
Dresden, Germany (2000).
\end{center}

\begin{center}
AD 2000 May 16
\end{center}

\section{Abstract}
The Ginibre ensemble of complex random Hamiltonian
matrices $H$ is considered.
Each quantum system described by $H$ is a dissipative system 
and the eigenenergies $Z_{i}$
of the Hamiltonian are complex-valued random variables.
For generic $N$-dimensional Ginibre ensemble
analytical formula for distribution of
second difference $\Delta^{1} Z_{i}$ of complex eigenenergies is presented.
The distributions of real and imaginary parts of
$\Delta^{1} Z_{i}$ and also of its
modulus and phase are provided for $N$=3.
The results are considered in view of Wigner and Dyson's
electrostatic analogy.
General law of homogenization of eigenergies for different
random matrix ensembles is formulated.

\section{Introduction}
\label{sec-introduction}

Random Matrix Theory RMT studies
quantum Hamiltonian operators $H$
for which their matrix elements
$H_{ij}$ are random variables
\cite{Haake 1990,Guhr 1998,Mehta 1990 0}.
There were studied among others the following
Gaussian Random Matrix ensembles GRME:
orthogonal GOE, unitary GUE, symplectic GSE,
as well as circular ensembles: orthogonal COE,
unitary CUE, and symplectic CSE.
The choice of ensemble is based on quantum symmetries
ascribed to $H$.
It was Eugene Wigner who firstly dealt with eigenenergy level repulsion
phenomenon \cite{Haake 1990,Guhr 1998,Mehta 1990 0}.
RMT is applicable now in many branches of physics:
nuclear physics (slow neutron resonances, highly excited complex nuclei),
condensed phase physics (fine metallic particles,  
random Ising model [spin glasses]),
quantum chaos (quantum billiards, quantum dots), 
disordered mesoscopic systems (transport phenomena).
Jean Ginibre considered another example of GRME
dropping the assumption of hermiticity of Hamiltonians
thus defining generic complex valued $H$
\cite{Haake 1990,Guhr 1998,Ginibre 1965,Mehta 1990 1}.
Hence, $H$ belong to general linear Lie group GL($N$, {\bf C}),
where $N$ is dimension and {\bf C} is complex numbers field.
Therefore, the eigenenergies $Z_{i}$ of quantum system 
ascribed to Ginibre ensemble are complex valued random variables.
Jean Ginibre postulated the following
joint probability density function 
of random vector of complex eigenvalues $Z_{1}, ..., Z_{N}$
for $N \times N$ Hamiltonian matrices 
\cite{Haake 1990,Guhr 1998,Ginibre 1965,Mehta 1990 1}:
\begin{eqnarray}
& & P(z_{1}, ..., z_{N})=
\label{Ginibre-joint-pdf-eigenvalues} \\
& & =\prod _{j=1}^{N} \frac{1}{\pi \cdot j!} \cdot
\prod _{i<j}^{N} \vert z_{i} - z_{j} \vert^{2} \cdot
\exp (- \sum _{j=1}^{N} \vert z_{j}\vert^{2}),
\nonumber
\end{eqnarray}
where $z_{i}$ are complex-valued sample points
($z_{i} \in {\bf C}$).
 
We emphasise here Wigner and Dyson's electrostatic analogy.
A Coulomb gas of $N$ unit charges moving on complex plane (Gauss's plane)
{\bf C} is considered. The vectors of positions
of charges are $z_{i}$ and potential energy of the system is:
\begin{equation}
U(z_{1}, ...,z_{N})=
- \sum_{i<j} \ln \vert z_{i} - z_{j} \vert
+ \frac{1}{2} \sum_{i} \vert z_{i}^{2} \vert. 
\label{Coulomb-potential-energy}
\end{equation}
If gas is in thermodynamical equilibrium at temperature
$T= \frac{1}{2 k_{B}}$ 
($\beta= \frac{1}{k_{B}T}=2$, $k_{B}$ is Boltzmann's constant),
then probability density function of vectors of positions is 
$P(z_{1}, ..., z_{N})$ Eq. (\ref{Ginibre-joint-pdf-eigenvalues}).
Therefore, complex eigenenergies $Z_{i}$ of quantum system 
are analogous to vectors of positions of charges of Coulomb gas.
Moreover, complex-valued spacings $\Delta^{1} Z_{i}$
of complex eigenenergies of quantum system:
\begin{equation}
\Delta^{1} Z_{i}=Z_{i+1}-Z_{i}, i=1, ..., (N-1),
\label{first-diff-def}
\end{equation}
are analogous to vectors of relative positions of electric charges.
Finally, complex-valued
second differences $\Delta^{2} Z_{i}$ of complex eigenenergies:
\begin{equation}
\Delta ^{2} Z_{i}=Z_{i+2} - 2Z_{i+1} + Z_{i}, i=1, ..., (N-2),
\label{Ginibre-second-difference-def}
\end{equation}
are analogous to
vectors of relative positions of vectors
of relative positions of electric charges.
The $\Delta ^{2} Z_{i}$ have their real parts
${\rm Re} \Delta ^{2} Z_{i}$,
and imaginary parts
${\rm Im} \Delta ^{2} Z_{i}$, 
as well as radii (moduli)
$\vert \Delta ^{2} Z_{i} \vert$,
and main arguments (angles) ${\rm Arg} \Delta ^{2} Z_{i}$.
$\Delta ^{2} Z_{i}$ are extensions of real valued second differences:
\begin{equation}
\Delta^{2} E_{i}=E_{i+2}-2E_{i+1}+E_{i}, i=1, ..., (N-2),
\label{second-diff-def}
\end{equation}
of adjacent ordered increasingly real valued energies $E_{i}$
defined for
GOE, GUE, GSE, and Poisson ensemble PE
(where Poisson ensemble is composed of uncorrelated
randomly distributed eigenenergies)
\cite{Duras 1996 PRE,Duras 1996 thesis,Duras 1999 Phys,Duras 1999 Nap}.

\section{Second Difference Distributions}
\label{sec-second-difference-pdf}
Let us define following random variables
for $N$=3 dimensional Ginibre ensemble:
\begin{equation}
Y_{1}=\Delta ^{2} Z_{1}, 
A_{1}= {\rm Re} Y_{1}, B_{1}= {\rm Im} Y_{1},
\label{Ginibre-Y1A1B1-def}
\end{equation}
\begin{equation}
R_{1} = \vert Y_{1} \vert, \Phi_{1}= {\rm Arg} Y_{1}.
\label{Ginibre-polar-second-diff-def}
\end{equation}
Their distributions for 
$N$=3 dimensional Ginibre ensemble read 
\cite{Duras 2000 JOptB}:
\begin{eqnarray}
& & f_{Y_{1}}(y_{1})=f_{(A_{1}, B_{1})}(a_{1}, b_{1})=
\label{Ginibre-marginal-pdf-Y1-def} \\
& & =\frac{1}{576 \pi} [ (a_{1}^{2} + b_{1}^{2})^{2} + 24]
\cdot \exp (- \frac{1}{6} (a_{1}^{2}+a_{2}^{2})).
\nonumber
\end{eqnarray}
\begin{equation}
f_{A_{1}}(a_{1})=
\frac{\sqrt{6}}{576 \sqrt{\pi}} (a_{1}^{4}+6a_{1}^{2}+ 51)
\cdot \exp (- \frac{1}{6} a_{1}^{2}),
\label{Ginibre-marginal-pdf-Y1Re-def}
\end{equation}
\begin{equation}
f_{B_{1}}(b_{1})=
\frac{\sqrt{6}}{576 \sqrt{\pi}} (b_{1}^{4}+6b_{1}^{2}+ 51)
\cdot \exp (- \frac{1}{6} b_{1}^{2}),
\label{Ginibre-marginal-pdf-Y1Im-def}
\end{equation}
\begin{eqnarray}
& & f_{R_{1}}(r_{1})=
\label{Ginibre-polar-second-diff-result} \\
& & \Theta(r_{1}) \frac{1}{288}r_{1}(r_{1}^{4}+24) \cdot \exp(- \frac{1}{6} r_{1}^{2}),
\nonumber \\
& & f_{\Phi_{1}}(\phi_{1})= \frac{1}{2 \pi}, \phi_{1} \in [0, 2 \pi].
\nonumber
\end{eqnarray}
For the generic $N$-dimensional Ginibre ensemble
we study complex-valued random variable of second difference:
$W_{1}=\Delta ^{2} Z_{1}$.
Its distribution is $P_{3}(w_{1})$ \cite{Duras 2000 JOptB}:
\begin{eqnarray}
& & P_{3}(w_{1})=
\label{W1-pdf-I-result} \\
& & = \pi^{-3} \sum_{j_{1}=0}^{N-1} \sum_{j_{2}=0}^{N-1} \sum_{j_{3}=0}^{N-1}
\frac{1}{j_{1}!j_{2}!j_{3}!}I_{j_{1}j_{2}j_{3}}(w_{1}),
\nonumber \\
& & I_{j_{1}j_{2}j_{3}}(w_{1})=
\label{W1-pdf-I-F} \\
& & = 2^{-2j_{2}} 
\frac{\partial^{j_{1}+j_{2}+j_{3}}}
{\partial^{j_{1}} \lambda_{1} \partial^{j_{2}} \lambda_{2}
\partial^{j_{3}} \lambda_{3}}
F(w_{1},\lambda_{1},\lambda_{2},\lambda_{3}) \vert _{\lambda_{i}=0},
\nonumber
\end{eqnarray}
\begin{eqnarray}
& & F(w_{1},\lambda_{1},\lambda_{2},\lambda_{3})=
\label{W1-pdf-I-F-final} \\
& & = A(\lambda_{1},\lambda_{2},\lambda_{3})
\exp[-B(\lambda_{1},\lambda_{2},\lambda_{3}) \vert w_{1} \vert^{2}],
\nonumber
\end{eqnarray}
\begin{eqnarray}
& & A(\lambda_{1},\lambda_{2},\lambda_{3})=
\label{W1-pdf-I-A} \\
& & =\frac{(2\pi)^{2}}
{(\lambda_{1}+\lambda_{2}-\frac{5}{4}) 
\cdot (\lambda_{1}+\lambda_{3}-\frac{5}{4})-(\lambda_{1}-1)^{2}},
\nonumber \\
& & B(\lambda_{1},\lambda_{2},\lambda_{3})=
\label{W1-pdf-I-B} \\
& & =(\lambda_{1}-1) \cdot \frac{2 \lambda_{1}-\lambda_{2}-\lambda_{3}+\frac{1}{2}}
{2 \lambda_{1}+\lambda_{2}+\lambda_{3}-\frac{9}{2}}.
\nonumber
\end{eqnarray}

\section{Conclusions}
\label{sect-conclusions}
In order to compare our results
for second differences for different ensembles 
we define dimensionless second differences:
\begin{equation}
C_{\beta} = \frac{\Delta^{2} E_{1}}{<S_{\beta}>},
\label{rescaled-second-diff-GOE-GUE-GSE-PE}
\end{equation}
\begin{equation}
X_{1}=\frac{A_{1}}{<R_{1}>},
\label{Ginibre-X1-def} 
\end{equation}
where $<S_{\beta}>$ are
the mean values of spacings 
for GOE(3) ($\beta=1$),
for GUE(3) ($\beta=2$),
for GSE(3) ($\beta=4$), for PE ($\beta=0$)
\cite{Duras 1996 PRE,Duras 1996 thesis,Duras 1999 Phys,Duras 1999 Nap},
and $<R_{1}>$ is mean value of radius $R_{1}$ 
for $N$=3 dimensional Ginibre ensemble \cite{Duras 2000 JOptB}.

We formulate homogenization law
\cite{Duras 1996 PRE,Duras 1996 thesis,Duras 1999 Phys,Duras 1999 Nap,Duras 2000 JOptB}: 
{\it Eigenenergies for Gaussian ensembles, for Poisson ensemble,
and for Ginibre ensemble tend to be homogeneously distributed.}
The second differences' distributions assume global maxima at origin
for above ensembles.
For Coulomb gas 
the vectors of relative positions of vectors
of relative positions of charges statistically tend to be zero.
It can be called stabilisation
of structure of system of electric charges.

\section{Acknowledgements}
\label{sect-acknowledgements}
It is my pleasure to most deeply thank Professor Jakub Zakrzewski
for formulating the problem.

\end{document}